\documentclass[11pt,twoside]{article}
\usepackage{asp2010}
\newcommand{\nz}{\ensuremath{\langle N_z\rangle}}

\newcommand{\te}{\ensuremath{T_{\rm eff}}}
\newcommand{\bz}{\ensuremath{\langle B_z \rangle}}

\resetcounters

\begin{document}

\title{Low magnetic fields in white dwarfs and their direct progenitors?}
\author{S. Jordan,$^{1}$ 
        S. Bagnulo,$^{2}$, 
          J. Landstreet,$^{2,3}$,
          L. Fossati ,$^{4}$,
          G.G. Valyanin,$^{5}$,
          D. Monin,$^{6}$,
          G.A. Wade,$^{7}$,
          K. Werner,$^{8}$,
          S.J. O'Toole,$^9$}
\affil{$^1$Astronomisches Rechen-Institut, Zentrum f\"ur Astronomie der Universit\"at Heidelberg, M\"onchhofstr. 12-14, D-69120 Heidelberg, Germany,
}
\affil{$^2$Armagh Observatory,
           College Hill,
           Armagh BT61 9DG,
           Northern Ireland, U.K.,}
\affil{$^3$
     Department of Physics and Astronomy, The University of
     Western Ontario, London, Ontario, N6A 3K7, Canada}
\affil{$^4$Argelander-Institut f\"{u}r Astronomie der Universit\"{a}t Bonn, 
     Auf dem H\"{u}gel 71, 53121 Bonn, Germany}
\affil{$^{5}$Special Astrophysical Observatory, Nizhnij Arkhyz, Zelenchukskiy
     Region, Karachai-Cherkessian Republic, Russia 369167}
\affil{$^{6}$Dominion Astrophysical Observatory, Herzberg Institute of
     Astrophysics, National Research Council of Canada, 5071 West
     Saanich Road, Victoria, BC V9E 2E7, Canada}
\affil{$^{7}$Department of Physics, Royal Military College of Canada,
     PO Box 17000, Stn Forces, Kingston, Ontario K7K 7B4, Canada}
\affil{$^{8}$Institute for Astronomy and Astrophysics, Kepler Center for Astro and
          Particle Physics, Eberhard Karls Universit\"at T\"ubingen, Sand 1, D-72076
T\"ubingen, Germany}
\affil{$^{9}$Australian Astronomical Observatory, P.O. Box 296, Epping, NSW
          1710, Australia}

\begin{abstract}
We  have carried out a re-analysis of polarimetric data of central stars of planetary nebulae, hot subdwarfs, and white dwarfs  taken with  FORS1  (FOcal
Reducer and
low dispersion Spectrograph) 
on the VLT (Very Large Telescope),  and added a large number of
new observations in order to increase the sample. A careful analysis of the observations using
only one wavelength calibration  for the polarimetrically analysed
  spectra and for all positions of the retarder plate of the spectrograph is crucial
in order to avoid spurious signals.  We find that the  previous detections of magnetic fields in subdwarfs and central stars could not be confirmed while
about 10\%\ of the observed white dwarfs  have magnetic fields
at the kilogauss level.
\end{abstract}

\section{Introduction}
Magnetic fields are found to occur in a wide variety of
stars, including pre-main sequence T Tau stars and Herbig AeBe stars,
upper main sequence O, B and A stars, rapidly rotating and active
lower main sequence stars, AGB stars, white dwarfs, and neutron
stars. Main sequence stars with effective temperatures below 7000\,K have
spatially complex magnetic fields and are thought to  be generated by current dynamos operating in the outer convective layer. 
Hotter stars
generally reveal fields in only a fraction of any stellar type, and
the fields appear simple in structure.  Such static fields are usually thought to be 
fossil fields, frozen into the star by the very high electrical
conductivity and originating from  earlier stages of the star's evolution.

Roughly 10\% of white dwarfs are found to host fields. Based
on modelling the magnetic effects in the optical spectrum, the fields
are found to range from tens of kG up to more than 1000~MG, with the
majority in the range of 1--100~MG.  

\citet{Kulebietal09}  have analyzed about 150 hydrogen-rich magnetic 
white dwarfs between 1 MG and 1200 MG. They found that only
about 50\%\ of all objects could be described by centred dipoles, while the other half shows clear
indications of offset dipoles 
(or higher-order modes). The analysis was performed by comparing the
 observed flux spectra
with models for the radiative transfer in the magnetised atmospheres
of white dwarfs using a least-squares method to find the
best-fitting field geometry, assuming magnetic dipoles offset relative
to the  centre of the star.

While for strong magnetic fields the magnetic field can be deduced by comparing both the flux and polarisation spectra  to theoretical models, the regime below
about 40\,kG can only be accessed by measurements of the circular polarisation with very large telescopes like the VLT. 
The study of available statistics by \citet{Liebertetal03}
 suggested that the detection
rate for field weaker than a few tens of kG may be significantly
higher than the frequency of $\sim 10$\,\%, which characterises the
overall detection rate of stronger fields \citep{Liebertetal2005}.

\citet{AznarCuadrado-etal:04,Jordan-etal:07}
have detected magnetic fields of  a few kG with the help of FORS1  
on the
VLT in three or four  (10\%) of the investigated objects.  
For each stellar observations, they obtained 4-14 integrations
with the quarter-wave plate rotated by $90^\circ$ between successive
exposures.

\begin{figure}[tp]
\centering
\includegraphics[width=1.0\textwidth]{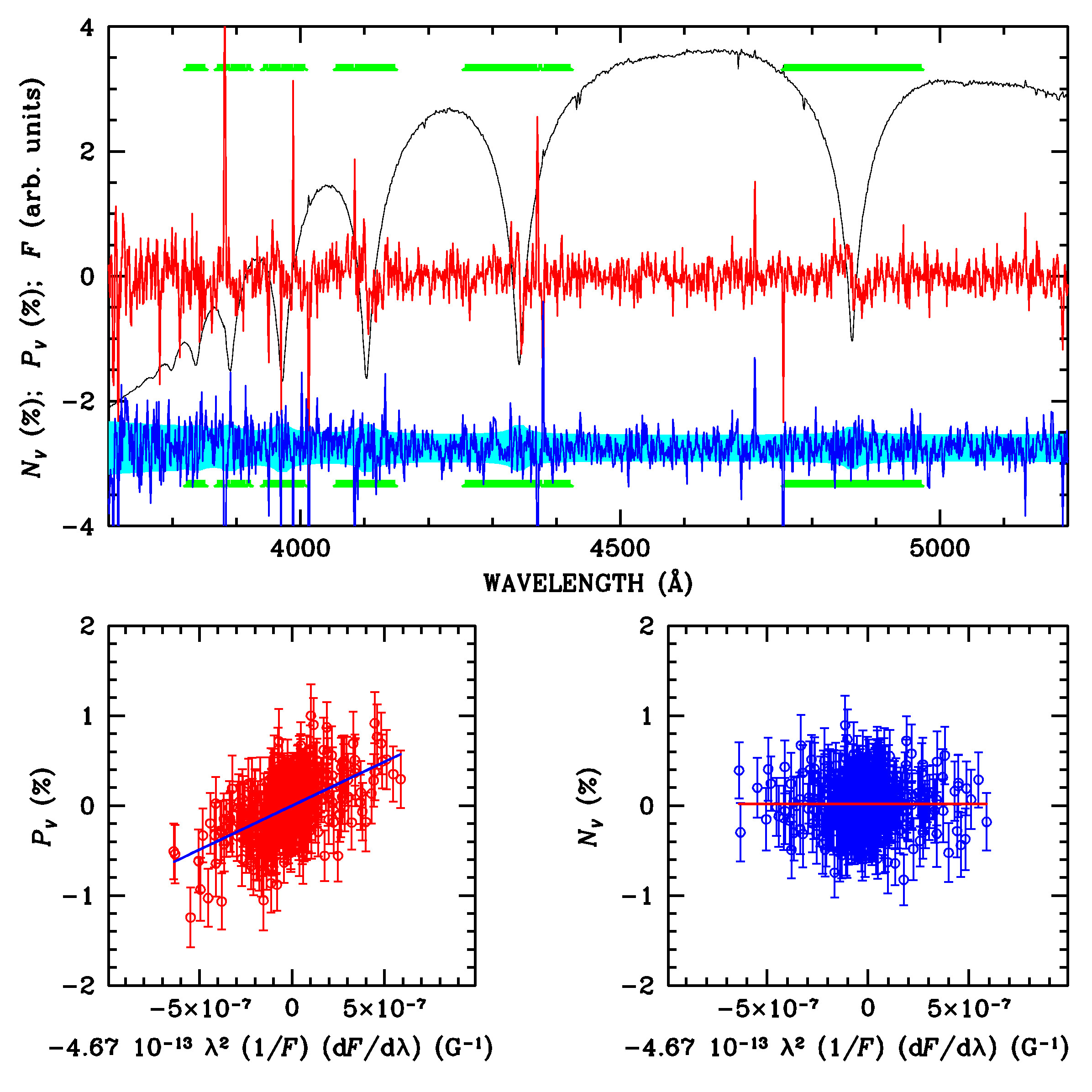}
\caption{\label{Fig:wd2105-820-5} 
Observations of WD2105-820 obtained with FORS1 on MJD 53 227.209. The top panel shows the observed flux $F$ (black solid line, in arbitrary
units), the circular polarisation profile $P_V = V/I$  (red solid line centred about 0), and the null profile $N_V$ (blue solid line, offset by  $-2.75$\% for
display purpose). The null profile, which is a test for the presence of spurious features, 
is expected to be centred about zero and scattered according to a Gaussian with $\sigma$ given by the $P_V$ error bars, which
are represented with light blue bars centred about $-2.75$\%. The regions used for field measurement are marked with green bars above and below this spectrum.
The slope of the interpolating lines in the bottom panels provides the mean longitudinal field from $P_V$ (left bottom panel) and from the null profile (right
bottom panel), both calculated using only the H Balmer lines. The corresponding \bz\ and \nz\ values are $9770 \pm 843$ ~G and $-11 \pm 868$~G, respectively.}

\end{figure}

The mean line-of-sight magnetic
field \bz\ was obtained by using the relationship
$V(\lambda) = -g_{\rm eff} C_{\rm Z} \lambda^2
\frac{{\rm d}I(\lambda)}{{\rm d}\lambda} \bz$
\citep{Landstreet82}, where $C_{\rm Z} = e/4\pi m c^2$.

The success of these observations led to similar surveys aiming at the
detection of kG magnetic fields in the direct progenitors of magnetic
white dwarfs,  namely
central stars of planetary nebulae and hot subdwarfs.

Hot subdwarfs are subluminous objects that dominate the population of faint
blue stars in our own galaxy. The spectral type sdB is defined by hydrogen-rich
atmospheres with effective temperatures below about 40\,000\,K
\citep[e.g.][]{Heber86}. The sdO
stars on the other hand cover a much larger range of atmospheric compositions
with a large spread of hydrogen and helium abundances; their effective
temperatures  range between 40000 and 90000\,K. In the
Hertzsprung-Russell diagram 
they are found  on the
extreme (blue end of the) horizontal branch.
The host subdwarfs  are believed to be the progeny of 1-2\%\ of the white dwarfs, the rest coming from
the central stars of planetary nebulae.

The question  of whether there are magnetic fields in central
stars of planetary nebulae is of particular importance
since there is still no conclusive theory capable of explaining why more than 80\,\%\ of  known planetary
nebulae (PNe) have bipolar and non-spherically symmetric structures
\citep{Zuckerman-Aller:86,Stanghellini-etal:93,CorradiSchwarz:95}. An overview
of the  mechanisms that may shape PNe is given by \cite{BalickFrank02}. Several
of these processes suggest that it is magnetic fields which deflect the
outflow of the matter along the magnetic field lines. 

 \cite{Jordan-etal:05} reported the
detection of magnetic fields in at least two  PNe,
 NGC\,1360 and LSS\,1362 with the same instrumentation used for the white dwarfs.
Even more surprising was the fact that \citet{OToole-etal:05} found kG magnetic fields in all four investigated hot subdwarf stars.

Recently, the detection of magnetic fields in the central stars of
planetary nebulae  was called into question by
\citet{Leone-etal:11}, who re-observed NGC\,1360 and LSS\,1362 with
the FORS2 instrument, and concluded that their effective magnetic
field is null within an uncertainty of $\sim 100$\,G (NGC\,1360), and
$\sim 290$\,G (LSS\,1362). Furthermore, both \citet{Leone-etal:11} and
\citet{Bagnulo-etal:12} re-analysed the observations previously
obtained with FORS1 by \citet{Jordan-etal:05}, and were unable to confirm
the original detection by Jordan et al.

\citet{Bagnulo-etal:12} analysed polarimetric observations from a large range of spectral types and found
that in many cases the results were spurious due to improper calibrations. \citet{Bagnulo-etal:09} have demonstrated
that it is very important to use one wavelength calibration 
  taken in one polarisation mode and for one position of the $\lambda/4$ retarder plate 
for all four combinations of polarisation mode and retarder plate orientation. In this case a number of sources of measurement errors are
cancelled out to first order. In fact it turned out that separate wavelength calibrations have been used instead so that spurious features leading
to false detection of magnetic fields could not be excluded.

For this reason all polarimetric VLT observations of central star of planetary nebulae, hot subdwarfs, and white dwarfs were
re-reduced with the state-of-the-art ESO FORS pipeline \citep{Izzo-etal:10} and additional observations from authors of this paper
were added to enlarge the samples.

\section{Results}
\subsection{Central stars of planetary nebulae}
We determined the magnetic fields from spectropolarimetric observations
of ten central stars of planetary nebulae. The results of the analysis included
the four stars investigated by \cite{Jordan-etal:05}  while the observations of six stars, plus additional
measurements of a star previously observed, were analysed for the first time.

All our determinations of magnetic field in the central planetary
nebulae were consistent with null results. Our field measurements
have a typical error bar of 150-300\,G. 
Therefore, we had to conclude that the  field   detections by \cite{Jordan-etal:05} were spurious.

For our sample of ten stars (Abell\,36 has been observed in both
observational campaigns), we conclude that there is no confirmed case of
a magnetic field in the central star of a planetary nebula at a kG
level. Magnetic fields of the order to 100-300\,G, however, cannot be
excluded. Indirect evidence of mG fields in
proto-planetary nebulae could still support an influence of magnetic
fields on the shape of PNe.

These results were published by \citet{Jordanetal12}.

\subsection{Hot subdwarfs}
After \citet{OToole-etal:05} detected field strengths of up to
$\sim$1.5\,kG range  at varying levels of significance in
 each of the  six targets stars, the question was whether these detections were real or whether 
the measurements were also spurious due to the application of a separate wavelength calibration to
each of the two polarisation modes and to each of the positions of the $\lambda/4$ retarder plate.

Therefore, we also repeated the reduction of these data and added new observations to clarify the question of how common magnetic fields are in subdwarf
stars. In total we have analysed a sample of 40 hot subdwarf stars of which 30 have been observed with the FORS1 and FORS2 instruments of the ESO VLT. 
It turned out that there is presently no strong evidence for the occurrence of a magnetic field in any sdB or sdO star, with typical longitudinal field
uncertainties of the order of 2-400\,G. 

The results of this investigation have been published by \citet{Landstreetetal2012a}.

\subsection{White dwarfs}
The negative outcome of the new investigations of the polarimetric data of central stars of planetary nebulae and hot subdwarfs has also cast doubt on the 
surveys aiming at the detection of kG magnetic fields in white dwarfs by \citet{AznarCuadrado-etal:04,Jordan-etal:07}.

In addition to re-reduction of the Aznar Cuadrado and Jordan data we have analysed  new observations of cooler 
 (DA6 -- DA8) white dwarfs, all taken with the  FORS1 spectrograph.

It turned out that some of the detections by
\citet{AznarCuadrado-etal:04,Jordan-etal:07} were  confirmed by
  the new reductions and 
that  longitudinal    magnetic fields weaker than 10\,kG have  been correctly
   identified in at least three white dwarfs. For one of
   these three weak-field stars (WD\,2359$-$434), UVES archive data
   show a $\sim 100$\,kG mean field modulus.  It could of course be that 
the  at the time of
   the FORS observations the star's magnetic field axis was nearly
   perpendicular to the line of sight, or the star's magnetic field
   has rather complex structure.  

 In addition, we  have discovered an apparently constant longitudinal magnetic
   field of $\approx 9.5$\,kG in the DA6 white dwarf WD\,2105$-$820.
   This star is the first weak-field white dwarf that has been
   observed sufficiently to roughly determine the characteristics of
   its field.  The available data are consistent with a simple dipolar
   morphology with magnetic axis nearly parallel to the rotation axis,
   and a polar strength of $\simeq 56$\,kG.  

In total we have now investigated 
20 hot DA stars (generally spectral type DA1 to DA4, $\te \ga 14000$\,K) and 15 cool
DA stars (spectral type DA5 to DA8; $\te \la 14000$\,K).
We  detected two magnetic white dwarfs in each of the hot and

lcool samples and conclude that detection rates are about 10\,\% for
the hot sample, and 13\,\% for the cool sample. The small size of the
sample and the small number of detections set a serious limit to
accuracy of these frequency estimates. Using the Wilson 95\,\%
confidence limits \citep{Wilson27}, the overall detection probability lies
between 4 and 25\%.

The field detection rate in hot
WDs could be anywhere between 2.8 and 30\,\%, while the field
detection rate in cool DA WDs lies between 3.7 and 38\,\%. 

Our data are consistent with the
hypothesis that weak magnetic fields occur with the same frequency in
hot and cool DA WDs.  Globally, the detection of four weak magnetic
fields from a total sample of 36 WDs makes it quite clear that the
probability of
detecting a $\sim 10$\,kG field in a WD is comparable to the
probability of detecting a magnetic field with strength in the range
100\,kG -- 500\,MG, which is $\sim 10$\,\%.

The results of this investigation have been published by \citet{Landstreetetal2012b}.

\section{Conclusion}
A re-analysis of the polarimetric measurements of central stars of planetary nebulae and  
hot subdwarfs as well as the analysis of newer observations have shown that there is no
evidence for the occurrence of kilogauss magnetic fields in these objects.

On the other hand magnetic fields  of a few kG have been
detected in about 10\%\ of all target  white dwarfs.

\acknowledgements 
We thank the staff of the ESO VLT for carrying out
  the service observations and providing support for the visitor mode observations.

\bibliography{jordan}

\end{document}